\documentstyle[12pt]{article}
\begin{document}
\baselineskip=18pt
\newcount\sectionnumber
\sectionnumber=0
\pagestyle{empty} 
\vspace{8mm}
\begin{flushright}
TAUP 2424-97\\
WIS-97/12/MAY-PH\\
UTPT-97-11
\end{flushright}
\begin{center}  {\bf {\huge  Hyperfine  Interactions in Charm and
Bottom Systems} }\\
\vspace{6mm}  
Harry J. Lipkin\\
Department of Particle Physics, \\
Weizmann Institute,\\
Rehovot 76100, Israel \\and\\
School of Physics and Astronomy, \\
Tel-Aviv University,\\ 
Tel-Aviv 69978, Israel \\

and

Patrick J. O'Donnell  \\
Physics  Department,\\  
University  of
Toronto,\\ 
Toronto, Ontario M5S 1A7, Canada.  
\end{center}

\vskip 20pt \centerline{\bf Abstract}

Hyperfine  interactions in the light meson and baryon sectors are
generalized  to the charm and bottom  systems.  It is pointed out
that an attempt to increase the value of the wave function at the
origin to account for the unusual ratio of  $\Lambda_{b}$  to the
$B^0$ lifetimes  could spoil the good agreement  among the baryon
and meson hyperfine  mass--splitting.  Including spin effects and
taking phase space  differences  into account we predict that the
decay rate of the $\Lambda_{b}$ can be increased relative to that
of the $B^0$ meson by about 7\%.

%\today

\newpage \pagestyle{plain}

\section{Introduction}

As the quark mass  becomes  heavier  many  differences  among the
properties of spin--$1/2$ and spin--$3/2$  baryons and also among
pseudoscalar  and  vector  mesons  containing  a heavy  quark are
expected  to  become  less  pronounced   \cite{Neubert}.  In  the
infinite mass limit there are model--independent  predictions for
all form factors in  transitions  from one heavy quark to another
in  terms  of  a  single   function  of  the  momentum   transfer
\cite{heavy}.  As the quark mass  increases  it is expected  that
the lifetimes of particles containing one heavy quark will become
very similar \cite{Bigi}.  It is in the corrections to the lowest
order in $\Lambda_{QCD}/M$ where models play a role.

Such was the case with the lifetimes of the $\Lambda_{b}$ and the
$B^0$  meson.  These  were  expected  to be the same in the heavy
quark  limit  and just  slightly  different  when  certain  quark
scattering  processes that could occur in the  $\Lambda_{b}$  but
not in the meson were included.  These  principally  included (a)
the  ``weak   scattering"   process,   first   invoked   for  the
$\Lambda_{c}^+$  lifetime  \cite{Barger},  and here of the  form,
$bu\rightarrow cd$, and (b), the so--called ``Pauli interference"
process  $bd\rightarrow  c\bar u dd$  \cite{Voloshin,Bilic}.  The
results of including  these terms is a slight  enhancement in the
decay rate  leading  to  $\tau(\Lambda_{b})/\tau(B^0)  \sim 0.9$,
whereas the evaluation \cite{pdg} of $\tau(\Lambda_{b})$ is $1.18
\pm 0.08$ ps and $\tau(B^0) = 1.56 \pm 0.06$ ps gives a very much
reduced fraction  $\tau(\Lambda_{b})/\tau(B^0)  = 0.73 \pm 0.06$,
or  conversely  a very much  enhanced  decay  rate.  (There  is a
recent CDF  result  \cite{CDF}  which  would  move this  fraction
higher  than the world  average  to a value of $0.85 \pm 0.10 \pm
0.05$).

The enhancement of the decay width, $\Delta  \Gamma(\Lambda_{b})$
from the $q - q$  scattering  involves  replacing  the usual flux
factor by  $|\psi(0)|^2$,  the wave function at the origin of the
pair of quarks $bu$ in the  $\Lambda_{b}$, (or the pair $bd$, for
which the wave  function is the same by isospin  symmetry).  This
wave  function  at the  origin  naturally  appears  in  hyperfine
splitting  \cite{dgg}.  Rosner \cite{Rosner} tried to account for
the enhancement by changing the wave function $|\psi(0)_{bu}|^2$; this would 
also correlate with the surprisingly large hyperfine splitting suggested by the 
DELPHI group \cite{DELPHI}.
He was able to show that, under certain  assumptions, there could
be at most a  $13\pm  7 $\%  increase  of the  amount  needed  to
explain the decay rate of the $\Lambda_{b}$.

In a more dramatic attempt to explain the lifetime problem it has
been    proposed    \cite{NS}   to   allow   the   ratio   $r   =
|\psi_{bq}^{\Lambda_b}(0)|^2/   |\psi_{b\bar  q}^{B_q}(0)|^2$  to
vary between $1/4$ and $4$.  Clearly such a large variation would
be ruled out by hyperfine relations.

Here we show that the  hyperfine  relations  among the mesons and
baryons  can be  generalized  to include  the  heavier  particles
leading to good  predictions  among the mass splittings.  We then
show that spin and  phase  space  effects  predict a  significant
enhancement of the $\Lambda_B$ decay rate of about 7\%.

\section{Hyperfine Interactions}

It is simplest to begin with the role of  hyperfine  interactions
in the meson  sector.  A number of years ago it was  pointed  out
\cite{FO}  that for the  ground  state  of  mesons,  $^3S_1$  and
$^1S_0$  (in  the  quark  model   spectroscopic   notation)   the
difference  $\delta  m^2 = m_{V}^2 -  m_{P}^2$  is  approximately
constant.  This holds very well for states that  contain at least
one light  quark and seems to be a  consequence  of the fact that
the quantity  $|\psi(0)_{q\bar  q}|^2/\mu_{ij}$, where $\mu_{ij}$
is the reduced mass of the $q - \bar q$ system, is  approximately
constant; this is an exact result if the confining potential is a
linear  one.  That  is,  for a  linear  confining  potential  the
quantity  $|\psi(0)_{q\bar  q}|^2/\mu_{ij}$ is independent of the
masses  of  the  quarks;  the   quantity   $(m_1  +m_2)   \langle
V_{hyp}\rangle$  is  independent of the flavors of the quarks and
the flavor  dependence of hadron wave functions.  This meant that
quantitative predictions for the hyperfine splitting could now be
obtained; due to differences in the wave functions these had been
difficult  to  give  .  The  empirical   regularity   that  meson
hyperfine splitting seem to be flavor independent when written as
differences  between the square of the masses was obtained.  From
this, a simple  formula in term of the quark  masses and a single
parameter  gave   predictions   for  the   $(^3S_1)_{s\bar   c}$,
$(^3S_1)_{u\bar  b}$, $(^3S_1)_{s\bar c}$ and $(^3S_1)_{s\bar b}$
mesons as well as the then known ones.

Thus we can write the basic physics of the hyperfine interactions
as
\begin{equation}
H = \sum_{i} m_i + V_{hyp}
\label{hyp1}
\end{equation}
with the expectation value of the hyperfine interaction being
\begin{equation}
\langle V_{hyp}\rangle = V \sum_{i>j} \frac{\langle {\bf s_i}.{\bf s_j}
\rangle}{m_i m_j} K_c |\psi(0)_{ij}|^2
\end{equation}
where $V$ is the  strength of the  hyperfine  interaction,  ${\bf
s_i}$ is the spin of the i-th  quark and $K_c$ is a color  factor
which is unity for a $q \bar q$ pair and $1/2$ for two  quarks in
a baryon.

To see how this empirical regularity  involving the difference of
the  squares of the masses  comes about for mesons  note that the
hyperfine interaction can be written as
\begin{equation}
\langle V_{hyp}\rangle = V \frac{\langle {\bf s_1}.{\bf s_2}\rangle}{m_1+ 
m_2}\frac{|\psi(0)_{12}|^2}{\mu_{12}}
\end{equation}
where  $\mu_{12}$  is the  reduced  mass of the  quark--antiquark
pair.  For the sum of the two  masses it is a good  approximation
to write
\begin{equation}
m_V+m_p \sim 2(m_1 + m_2)
\end{equation}
giving
\begin{equation}
m_{V}^2 - m_{P}^2  \sim 2(m_1 + m_2)\langle V_{hyp}\rangle
\end{equation}

One of us  \cite{Lipkin}  generalized  this  result to baryons by
applying the identity used in atomic physics  \cite{Hiller} which
relates the wave function at the origin to the  derivative of the
potential.  If $K_c  W_{12}$  denotes  the  two--body  attractive
potential  responsible  for the quarks being bound, then, for the
meson system,
\begin{equation}
\frac{|\psi(0)_{12}|^2}{K_c \mu_{12}} = \langle \frac{dW_{12}}{dr_{12}}
\rangle
\label{potential}
\end{equation}

For  baryons,  $W_{12}$ in Eq.  (\ref{potential})  is replaced by
the total potential for the three--body system and the derivative
is   that  of  the   relative   coordinate   $r_{12}$.  For   the
non--strange  baryons,  with a  totally  symmetric  spatial  wave
function, this was used to obtain the constraint \cite{Lipkin},
\begin{equation}
\frac{|\psi(0)_{12}|^2}{K_C \mu_{12}} = \langle 
\frac{dW_{12}}{dr_{12}}\rangle 
(17/12 \pm 1/12)
\end{equation}
or, for spin--$3/2$ baryons and spin--$1$ mesons
\begin{equation}
<V_{hyp}>_{B(3/2)} = (17/16 \pm 1/16)\frac{q_M}{q_{B}}<V_{hyp}>{M(1)}
\end{equation}
where $q_M$ and $q_B$ are effective quark masses in the meson and
baryon.  This led to the relation
\begin{eqnarray}
M_{\Delta} - M_{N} &=& (17/32 \pm 1/32)\frac{q_M}{q_{B}}(M_{\rho} - M_{\pi})  
\nonumber\\
293 {\rm MeV} & = & 279 \pm 16 {\rm MeV}.
\end{eqnarray}

For the strange hyperons, there is no overall spatial permutation
symmetry  and the quark  masses are not equal, a situation  which
becomes  even more obvious in both the charm and bottom  systems.
However, as the non  equal--mass  quark becomes  heavier we might
expect  the  approximation  used  \cite{Lipkin}  of  setting  the
effective  quark masses in the meson and baryon to be equal to be
more  exact.  Incidentally,  this  approximation  gives  the mass
relations for the strange quark systems,
\begin{eqnarray}
M_{\Sigma^{*}} - M_{\Sigma} \qquad &= M_{\Xi^{*}} - M_{\Xi} \qquad
 &=  (17/32 \pm 
1/32)\frac{q_M}{q_{B}}(M_{K^*} - M_{K})  \nonumber\\
193.43 \pm 0.06 {\rm MeV} &= 213.68 \pm 0.086 {\rm MeV} &= 211 \pm 12 {\rm 
MeV} .
\end{eqnarray}

The  prediction  is  reasonably  good even though it ignores wave
function   effects  in  the  strange   quark  system  of  baryons
\cite{Cohen}.  As we get into the  heavy  quark  regime  we might
expect the analogous relations to be better.

A check on the charm  sector  seems to bear this out, for, in the
same type of approximation we have the relation,
\begin{eqnarray}
M_{\Sigma_{c}^{*}} - M_{\Sigma_{c}} \quad &=  M_{\Xi_{c}^{*}} - M_{\Xi_{c}^s} 
\quad &=(17/32 \pm 
1/32)\frac{q_M}{q_{B}}(M_{D^{*}} - M_{D}) \nonumber\\
65.7 \pm 0.06 {\rm MeV}&=63.2 \pm 2.6 {\rm MeV}&= 75 \pm 4.4 {\rm MeV} .
\end{eqnarray}

Here we have used the recent precise  measurement for the mass of
$\Sigma_{c}^{*}$ reported by the CLEO collaboration  \cite{CLEO}.
The   middle   number   comes   from   a   theoretical   estimate
\cite{Jenkins}  based on an  expansion  in  $1/m_Q$,  $1/N_c$ and
SU(3)  flavor  breaking.  There is no  measurement  as yet of the
mass of the  $\Xi_{c}^s$.  If the wave  function of the $bu$ pair
in the baryon is enhanced using the  prescription of Rosner then,
in the charm  sector, we would  actually  have a decrease  in the
wave  function  of  the  $cu$  pair  relative  to  that  for  the
$q$--$\bar q$ pair in the mesons.

Since the attempt at enhancing the wave  function did not succeed
in  explaining  the  $\Lambda_{b}$  lifetime, and even leads to a
decrease for the charm  sector it may be that the older  symmetry
results  \cite{FO,  Lipkin}  should not be abandoned,  especially
since they are also capable of  accounting  for the ratios of the
decay constants  \cite{me};  $f_B/f_D \sim 0.63$  \cite{me} to be
compared  with the  value  $0.79  \pm  0.21$  deduced  by  Rosner
\cite{Rosner}.  We now apply these hyperfine relations to the $b$
system.  There  has been a  reported  measurement  by the  DELPHI
collaboration  \cite{DELPHI}  of the large  value $56 \pm  16{\rm
MeV}$   for   the   mass   difference    $M_{\Sigma_{b}^{*}}    -
M_{\Sigma_{b}}$.  If we use this to check the  comparison  we end
up with

\begin{eqnarray}
M_{\Sigma_{b}^{*}} - M_{\Sigma_{b}} &=&  (17/32 \pm 
1/32)\frac{q_M}{q_{B}}(M_{B^{*}} - M_{B}) \nonumber\\
56 \pm 16 {\rm MeV}&=& 24.4 \pm 1.4 {\rm MeV} .
\label{b}
\end{eqnarray}
which is out of line with all of the preceding  comparisons.  (No
new  analysis has been done since the  conference  report in 1995
\cite{Feindt}).  This smaller  value on the right hand side of Eq.
(\ref{b})  is more in line  with a recent  update  of the  baryon
masses based on an expansion in $1/m_Q$, $1/N_c$ and SU(3) flavor
breaking\cite{Jenkins},  which  predicts  an  even  smaller  mass
difference  of $15.8  \pm  3.3$  MeV,  with the  errors  being an
estimate of the  uncertainty in scaling up from the charm sector.
The  value  of $56$  MeV was  part  of the  motivation  for a new
description \cite{Falk} of the heavy baryons.

\section{ Effects of hyperfine splitting on phase space }

We now note  another  way  that  spin  effects  can  enhance  the
$\Lambda_b$ lifetime relative to the $B$ lifetime.

Consider an extreme factorization assumption in which the b-quark
decay is  described  for any given  exclusive  decay mode both in
meson and baryon decays by
\begin{equation}                                                                
b \rightarrow c  +  X
\label{spect1}    
\end{equation}                        								
where  $X$  denotes  any  hadron  or  multihadron  state  and the
hadronic  decay is described by combining the charmed  quark with
the  spectator  antiquark  in the B decay and with the  spectator
diquark in the  $\Lambda_b$  decay.  We further  assume  that the
spins  of  the  spectator  quarks  are  not  changed  during  the
transition.

We then find that the following  hadronic  decays are relevant if
the c-quark  combines  with the  spectators  in the ground  state
configuration,
\begin{eqnarray}                                                           
B \rightarrow D~  +  X                 
\nonumber\\                                                                     
B \rightarrow D^* + X                         
\nonumber\\                                                                     
\Lambda_b \rightarrow \Lambda_c + X           
\end{eqnarray}                                                                  

The  assumption of no change in spectator  spin is not crucial to this  
argument; the spin of the spectator  diquark may be flipped by gluon exchange  
between the charmed quark and the diquark, but the isospin cannot be changed  
because the state has isospin zero and cannot be changed by strong QCD 
interactions when the charmed quark combines with the diquark.  
The  argument  should be  checked  experimentally,  as  discussed
below,  as soon as data  on  $\Lambda_b$  decays  into  different
baryon  final  states  are  available.  This   ``spectator   spin
conservation"  leads  to spin  factors  that  favor  the  baryons
because  the two quarks in the  $\Lambda_b$  are  coupled to spin
zero  and  can  only  combine  with  the  $c$  quark  to  make  a
$\Lambda_c$ and not a $\Sigma_c$ or $\Sigma^*_c$, while the spins
of the c quark and the spectator  antiquark are  uncorrelated  in
the $B$  decay and favor  the  $D^*$  over the $D$ by a factor of
3:1.  The  resulting  spectrum of final  states in the meson case
has the hyperfine  energy  averaged out, while in the baryon case
our spectator assumption chooses the final state in the multiplet
with the lowest hyperfine  energy.  The added hyperfine energy is
available for the  transition and leads to an enhancement  in the
phase space for the baryon  transition over the meson transition. (A different 
argument \cite{Altarelli} using the scaling of lifetimes as the inverse fifth 
power of hadronic rather than quark masses implicitly gives a larger phase 
space also.)

A very rough calculation shows this hyperfine energy  enhancement
of the baryon phase space.  A proper  calculation  would choose a
mass for X and  calculate  the  relative  momentum  for the three
transitions,   taking  also  into   account  the  effect  of  the
$D-\Lambda_c$  mass  difference  on the recoil  energy which also
favors the baryon  transition.  One might take the mass of $X$ to
be 1 GeV for a nonstrange-noncharmed transition and 2 GeV for the
case where the $W$ turns into a $D_s$.

Present  B-decay data support this set of assumptions but are not
sufficiently  precise to be convincing.  Decays into final states
containing  $D$ and vector $D^*$ modes are  observed  while those
containing  higher $D^*$'s are not.  This supports the assumption
that  only  the  ground  state  configurations  give  appreciable
contributions  to phase space.  The  semileptonic  partial widths
show these spin factors in the final states clearly since the 3:1
factor favoring the $D^*$ over the $D$ seems to be present.

Unfortunately  there are not  enough  data  about  the  exclusive
branching ratios, particularly for the baryons, to say more.

Another way to express  this spin effect is to assume  that the B
and $\Lambda_b$  decays would have the same phase space if hadron
spectroscopy  could be ignored, and the decays would sum over all
final  states   without   regard  to  spin  form   factors.  This
assumption is violated by spectator spin  conservation,  since it
requires the u and d pair in the initial $\Lambda_b$ to remain in
a spin-zero  state in the final state and not in  spin-one.  This
immediately  leads to a number of interesting  predictions  which
can be checked  by future  experiments.  The  $\Lambda_{b}$  will
decay to a $\Lambda_c$ and not a $\Sigma_c$ or $\Sigma_c$* if the
spin-zero  diquark picks up a charmed  quark, to a $\Lambda$  and
not a $\Sigma$ or $\Sigma^*$ if it picks up a strange  quark, and
to a  nucleon  and not a  $\Delta$  if it  picks up a  nonstrange
quark.

For a  quantitative  estimate we use the following  toy model for
semileptonic decays:  We assume that the $\Lambda_b$ goes only to
$\Lambda_c$, that the $B$ goes to a statistical  mixture (3/4) D*
and  (1/4) D and  that  all  transitions  to  higher  states  are
negligible.  The phase  space for the  $\Lambda_b$  decay is then
given by the mass difference $\Lambda_b - \Lambda_c$ to the fifth
power.  The  phase  space for the $B$ decay is then  given by the
$B-D^*$  mass  difference  to  the  fifth  power,  weighted  by a
statistical factor of (3/4) plus the $B-D$ mass difference to the
fifth power, weighted by a statistical factor of (1/4).

This well-defined  model for semileptonic  decays may be right or
wrong, but its  predictions  are easily  calculated and the basic
assumptions can be easily tested when exclusive  branching ratios
into baryon final states  including  spin-excited  baryons become
available.  We  immediately  obtain the following  result for the
ratio of semileptonic partial widths:
\begin{equation}                                                                
{{\Gamma (\Lambda_b)}\over {\Gamma(B)}} =1.07 
\label{spectwidth} 
\end{equation}  
In a toy model where this is the  dominant  decay mode this would
give the ratio of the lifetimes
\begin{equation}                                                                
{{\tau(\Lambda_b)}\over {\tau(B)}} =0.938        
\label{specttime} 
\end{equation}                          

This shows a clear prediction of a significant enhancement of the
$\Lambda_b$  partial  semileptonic  width in comparison  with the
$B$.  The $\Lambda_b$ decay rate is enhanced by about 7\%.

These  results  suggest  (1) that  phase  space  effects  must be
carefully  taken into account  using  exclusive  final  states in
lifetime  calculations  which  compare  the  $B$ and  $\Lambda_b$
decays; (2) that the validity of the spectator spin  conservation
model  should  be tested  with new data and new  analyzes  to see
whether the $\Lambda_b$ decay branching ratios show the predicted
predominance of  $\Lambda_c$,  $\Lambda$ and nucleon in exclusive
final  states in  comparison  with  $\Sigma_c$  and  $\Sigma_c$*,
$\Sigma$ and $\Sigma^*$, and $\Delta$, respectively . If only isospin is 
conserved, and not spectator spin, the $\Sigma_c$ and $\Sigma^{*}_{c}$ can  
appear only if accompanied by an appropriately charged $\pi$ coming from the 
decay of a higher isoscalar resonance. Thus, in a decay $\Lambda_{b} 
\rightarrow \Lambda_{c}$, there should be two oppositely charged pions with the 
following mass constraints; $m_{\Lambda_{c}} + m_{\pi} = m_{\Sigma^{(*)}_{c}}$ 
while the  mass of the $\Lambda_{c}$ together with the mass of the two pions 
should add up to the mass of an isoscalar resonance.

\vspace{.3in}

\centerline{ {\bf  Acknowledgment}}

This work was  supported  in part by grant No.  I-0304-120-.07/93
from The  German-Israeli  Foundation for Scientific  Research and
Development and by the Natural  Sciences and Engineering  Council
of Canada.  We thank J. Appel and Salam Tawfiq for helpful comments.


\newpage

\begin{thebibliography}{99}
\bibitem{Neubert} For a review see M. Neubert, Phys. Rep. {\bf 245}, 259 
(1994).
\bibitem{heavy}  N.  Isgur and M.  B.  Wise, Phys.  Lett.  B {\bf
232}  (1989)  113; {\sl ibid} B {\bf 237}  (1990)  527; M.  Luke,
Phys.  Lett.  B {\bf 252} (1990)  447; M.  Neubert,  Phys.  Lett.
B  {\bf  341}  (1995)  367;  I.  I.  Bigi,  M.  Shifman,   N.  G.
Uraltsev, and A.  Vainshtein, Phys.  Rev.  Lett.  {\bf 71} (1993)
496; A.  Manohar and M.  B.  Wise, Phys.  Rev.  D {\bf 49} (1994)
1310; M.  Luke and M.  Savage,  Phys.  Lett.  B {\bf 321}  (1994) 88.
\bibitem{Bigi} A recent review is by I.Bigi, M. Shifman and N. Uraltsev, 
hep-ph/9703290.
\bibitem{Barger}  V. Barger, J. P. Leveille and P. M. Stevenson, Phys. Rev. 
Lett. {\bf 44}, 226 (1980).
\bibitem{Voloshin} M . B. Voloshin and M. A. Shifman, Yad. Fiz. {\bf 41} 187 
(1985) [Sov. J. Nucl. Phys. {\bf 41}, 120 (1985)]; Zh. Eksp. Teor. Fiz {\bf 
91}, 1180 (1986) [Sov. Phys. - JETP {\bf 64}, 698 (1986)]; M . B. Voloshin, N. 
G. Uraltsev, V. A. Khoze and M. A. Shifman, Yad. Fiz. {\bf 46}, 181 (1987) 
[Sov. J. Nucl. Phys. {\bf 46}, 112 (1987)].
\bibitem{Bilic} N. Bili\'{c}, B. Guberina and J. Trampeti\'{c}, Nucl. Phys. 
{\bf B248}, 
261 (1984); B. Guberina, R. Ruckl and J. Trampeti\'{c}, Zeit. Phys {\bf C33}, 
297 
(1986).
\bibitem{pdg} R. M. Barnett et. al., Phys. Rev {\bf D54}, 1 (1996).
\bibitem{CDF} J. Tseng, Fermilab preprint Fermilab-Conf-96/438-E, to appear in 
the proceedings of the Second International Conference on Hyperons, Charm and 
Beauty Hadrons, 1996.
\bibitem{dgg} A. De Rujula, H. Georgi and S. L Glashow, Phys. Rev. {\bf D12}, 
147 (1975).
\bibitem{Rosner} J. L. Rosner, Phys. Lett {\bf B379}, 267 (1996).
\bibitem{DELPHI} DELPHI Collaboration, DELPHI 95-107.
\bibitem{NS} M. Neubert and C.T. Sachrajda, Nucl. Phys. {\bf B 483}, 339 
(1997); M. Neubert, talk at B Physics Conference, Hawaii, March 1997.
\bibitem{FO} M. Frank and P. J. O'Donnell, Phys. Lett. {\bf B159}, 174 
(1985); Zeit. Phys. {\bf C34}, 39 (1987).
\bibitem{Lipkin} H. J. Lipkin, Phys. Lett. {\bf B171}, 293 (1986); Phys. Lett 
{\bf B172}, 242 (1986).
\bibitem{Hiller} J. Hiller, J. Sucher and G. Feinberg, Phys. Rev {\bf A18}, 
2399 (1978).
\bibitem{CLEO} CLEO Collaboration, G. Brandenburg {\it et. al.}, 
\bibitem{Cohen} I. Cohen and H. J. Lipkin, Phys. Lett. {\bf B106}, 119 (1981).
\bibitem{Jenkins} E. Jenkins Phys. Rev. {\bf D 55}, R10 (1997). See also, M. 
Savage, Phys. Lett {\bf B359}, 189 (1995).
\bibitem{me} P. J. O'Donnell Phys. Lett {\bf B261}, 136 (1991).
\bibitem{Feindt} M. Feindt, private communication.
\bibitem{Falk} Adam F. Falk, Phys. Rev. Lett. {\bf 77}, 223 (1996).
\bibitem{Altarelli} G. Altarelli {\it et. al.}, Phys. Lett. {\bf B382}, 409 
(1996).
\end{thebibliography}
\end{document}